# Rotation symmetry breaking in the normal state of a kagome superconductor $KV_3Sb_5$

Hong Li[1,*], He Zhao[1,*], Brenden R. Ortiz[2], Takamori Park[3], Mengxing Ye[4], Leon Balents[4], Ziqiang Wang[1], Stephen D. Wilson[2] and Ilija Zeljkovic[1]

[1] *Department of Physics, Boston College, Chestnut Hill, MA 02467, USA*

[2] *Materials Department and California Nanosystems Institute, University of California Santa Barbara, Santa Barbara, California 93106, USA*

[3] *Department of Physics, University of California Santa Barbara, Santa Barbara, California 93106, USA*

[4] *Kavli Institute for Theoretical Physics, University of California, Santa Barbara, Santa Barbara, California 93106, USA*

*\* equal contribution*

*Correspondence:* ilija.zeljkovic@bc.edu

**Recently discovered kagome superconductors $AV_3Sb_5$ ($A$=K, Rb, Cs) provide a fresh opportunity to realize and study correlation-driven electronic phenomena on a kagome lattice. The observation of a $2a_0$ by $2a_0$ charge density wave (CDW) in the normal state of all members of $AV_3Sb_5$ kagome family has generated an enormous amount of interest, in an effort to uncover the nature of this CDW state, and identify any "hidden" broken symmetries. We use spectroscopic-imaging scanning tunneling microscopy to reveal a pronounced intensity anisotropy between different $2a_0$ CDW directions in $KV_3Sb_5$. In particular, by examining the strength of ordering wave vectors as a function of energy in Fourier transforms of differential conductance maps, we find that one of the CDW directions is distinctly different compared to the other two. This observation points towards an intrinsic rotation symmetry broken electronic ground state, where the symmetry is reduced from $C_6$ to $C_2$. Furthermore, in contrast to previous reports, we find that the CDW phase is insensitive to magnetic field direction, regardless of the presence or absence of atomic defects. Our experiments, combined with earlier observations of a stripe $4a_0$ charge ordering in $CsV_3Sb_5$, establish correlation-driven rotation symmetry breaking as a unifying feature of $AV_3Sb_5$ kagome superconductors.**

Quantum materials built from atoms arranged on a kagome network are predicted to exhibit unconventional electronic behavior [1–7] due to non-trivial Berry phase effects and strong electronic interactions. Experimental efforts on this front have primarily focused on transition-metal kagome magnets [8–16], in pursuit of realizing exotic phenomena such as topological flat bands, Weyl nodes and tunable Dirac fermions. Distinct from magnetically-ordered kagome systems, recently discovered kagome metals $A$V$_3$Sb$_5$ [17] ($A$=K, Rb, Cs) do not exhibit a resolvable magnetic order [17–19]. Surprisingly however, they show a large unconventional anomalous Hall response [20,21]. Furthermore, in striking resemblance to phenomena observed in high-$T_c$ superconductors, $A$V$_3$Sb$_5$ also exhibit various density wave phases [18,20,22–30] and superconductivity [18,24–26,28,29,31–36], including a unidirectional $4a_0$ charge order [24,26] and the potential emergence of a Cooper pair density wave [26] in CsV$_3$Sb$_5$.

Aside from superconductivity, a common feature identified across all members of this kagome family is a $2a_0$ x $2a_0$ charge density wave (CDW) that develops in the normal state above the superconducting transition ($T_{CDW} \approx$ 80-100 K [18,27,30,33]). This CDW phase has attracted a tremendous amount of experimental [18,20,22–30,37] and theoretical interest [38–43], as the first symmetry-broken phase that emerges upon cooling the system down from room temperature. As such, other phases that form at lower temperatures develop from this symmetry-broken state, and may carry the fingerprint of the $2a_0$ CDW phase. The majority of theoretical proposals and experiments to-date focused on modeling and understanding the $2a_0$ by $2a_0$ CDW as an electronic phase where the three CDW directions are identical (3Q-CDW). In this work, using spectroscopic-imaging scanning tunneling microscopy (SI-STM), we reveal an intrinsic rotation symmetry breaking in the $2a_0$ x $2a_0$ CDW phase of KV$_3$Sb$_5$. By exploring the amplitude of CDW peaks in Fourier transforms of STM differential conductance maps as a function of energy, we find that one of the CDW directions is markedly different than the other two, which reduces the rotation symmetry from $C_6$ to $C_2$. By imaging the same area of the sample with the same STM tip in different magnetic fields, we reveal that this directionality is not sensitive to the magnitude or the direction of magnetic field, which is in contrast to a recent report of an unconventional CDW tunable by magnetic field in KV$_3$Sb$_5$ [23]. Our work establishes a unifying picture of rotation symmetry breaking as a generic feature of $A$V$_3$Sb$_5$ kagome superconductors.

KV$_3$Sb$_5$ is a layered kagome superconductor ($T_c \approx$ 0.9 K) characterized by a hexagonal crystal structure ($a = b$ = 5.4 Å, $c$ = 9 Å) [33] composed of alternately stacked V-Sb slabs and K layers (Fig. 1a,b). Each V-Sb slab can be described by a kagome lattice of V atoms, interweaved by a hexagonal lattice of Sb atoms. Consistent with previous work [23–26], we find that the sample cleaves between the alkali layer (K) and the Sb layer. This exposes either a K surface often prone to reconstruction (Supplementary Figure 1), or a complete Sb surface, which we focus on in this work. STM topographs of the Sb layer show a honeycomb-like surface structure (Fig. 1c). To visualize the large-scale electronic band structure, we use quasiparticle interference (QPI) imaging. This method relies on the detection of elastic scattering and interference of electrons as static, periodic charge modulations in differential conductance d$I$/d$V$(**r**,$V$) maps. In our sample, the FTs of d$I$/d$V$(**r**,$V$) maps show an isotropic scattering vector **q**$_1$ near the FT center in momentum-transfer space (**q**-space) (Fig. 2a,b). Similarly to the spectroscopic mapping of the Sb surface of CsV$_3$Sb$_5$ [24], **q**$_1$ can be observed across a wide range of energies. Its magnitude and dispersion is consistent with an electron-like band at the center of the Brillouin zone, primarily associated with Sb orbitals (Fig. 2c). The Fermi vector **k**$_f \approx$ 0.18 Å$^{-1}$ determined from our data (**q**$_1$($E$ = 0) = 2**k**$_f$) also shows a close match to that measured by angle-resolved photoemission spectroscopy [21] and expected from calculations [33], thus demonstrating an approximate consistency between theory and different measurement techniques.

In addition to the hexagonal lattice, STM topographs also reveal periodic conductance variations between neighboring unit cells (Fig. 1c,d). This is consistent with the established $2a_0 \times 2a_0$ CDW phase of $A$V$_3$Sb$_5$ [18,20,23–29], with momentum-space ordering wave vectors $\mathbf{Q}_{2a0}^i = \frac{1}{2}\mathbf{Q}_{Bragg}^i$ ($i=a, b, c$) that can be clearly seen in the FT of an STM topograph (Fig. 1d). We note that the FT peak corresponding to a $4a_0$ charge ordering peak observed in CsV$_3$Sb$_5$ [24,26] is absent in KV$_3$Sb$_5$ (Fig. 1d). To demonstrate the CDW origin of $\mathbf{Q}_{2a0}^i$ peaks, as opposed to energy-dispersive QPI features, we point that these peaks do not disperse in FTs of d$I$/d$V$($\mathbf{r}$,$V$) maps as a function of bias (Fig. 2d). We note that we apply the Lawler-Fujita drift-correction algorithm [44] to all our data to align the atomic Bragg peaks onto single pixels with coordinates that are even integers (defined with respect to the center of the FT). This processing method in turn makes the CDW FT peaks $\mathbf{Q}_{2a0}^i = \frac{1}{2}\mathbf{Q}_{Bragg}^i$ ($i =a, b, c$) also confined to a single pixel, which enables an easy read-out and comparison of CDW amplitudes at different wave vectors and across different data sets. This process also minimizes the smearing of the FT peaks across neighboring pixels due to a small piezoelectric drift and thermal effects. Our first observation is that different CDW peaks display a small difference in amplitude (peak height), as shown in Fig. 1d. Moreover, the relative amplitude between different peaks depends on the imaging bias, and the anisotropy between different directions becomes very pronounced near Fermi level (Fig. 1e). Although STM tips are typically somewhat anisotropic, we point that the amplitude of the strongest atomic Bragg peak $\mathbf{Q}_{Bragg}^b$ actually corresponds to the weakest CDW peak $\mathbf{Q}_{2a0}^b$ (inset in Fig. 1d). As we will demonstrate below, this directionality is not a trivial consequence of the STM tip anisotropy and it is rooted in the underlying rotation symmetry breaking of the electronic structure.

To explore this further, we track the evolution of the CDW peak amplitudes in the FTs of d$I$/d$V$($\mathbf{r}$,$V$) maps (Fig. 3a-c). For a conventional 3Q-CDW phase, the three CDW directions $\mathbf{Q}_{2a0}^i$ should be equivalent, and as such, the amplitudes at the corresponding FT wave vectors should in principle follow the same energy dependence. In contrast to this, we find that the peak amplitude along one of the CDW directions exhibits a markedly different energy dependence compared to the other two. For example, in Fig. 3c, we can observe that the amplitude profile associated with $\mathbf{Q}_{2a0}^a$ and $\mathbf{Q}_{2a0}^c$ is nearly identical, but the one related to $\mathbf{Q}_{2a0}^b$ is noticeably different. This results in pronounced unidirectionality in d$I$/d$V$($\mathbf{r}$,$V$) maps that can also be observed in real-space (Fig. 3d). The slight difference between the two equivalent peaks $\mathbf{Q}_{2a0}^a$ and $\mathbf{Q}_{2a0}^c$ may be related to a small tip anisotropy. Understanding the amplitude profiles of different peaks in Fig. 3d is beyond the scope of this work, and we note they can appear differently for different samples and tips. For instance, in Fig. 4e we can observe that amplitude dispersions along two CDW directions result in a sharp peak around 5 mV, while in Fig. 3d only one amplitude dispersion exhibits a peak at that same energy. Nevertheless, our observation that one CDW peak differs from the other two is robust across multiple samples scanned with different STM tip wires (Fig. 3c, Fig. 4e). This provides strong evidence for that the electronic ground state is C$_2$-symmetric at low temperature.

The experiments discussed thus far have all been performed at zero external magnetic field. Given the proposals of time-reversal symmetry-breaking orbital currents [38,40] associated with the CDW phase, it is crucial to determine if and how the observed CDW signal couples to external magnetic field [23]. To investigate this, we repeat the STM measurements as a function of magnetic field $B$. The relative heights of all CDW peaks in FTs of STM topographs appear comparable before and after a moderate external field is applied perpendicular to the $c$-axis (Fig. 4d). Importantly, the amplitudes are insensitive to the magnetic field direction (i.e. $B$ applied parallel and antiparallel to the $c$-axis) (Fig. 4d). The robustness of CDW to magnetic field direction reversal is in contrast to a recent report of a CDW order tunable by magnetic field in the same material [23]. We note that the area imaged in Fig. 4 does not contain defects occasionally seen in for example Fig. 2a, so defect pinning may not explain the contrasting observations, as hypothesized in Ref. 23. The observations reported here are confirmed in bias-dependent d$I$/d$V(\mathbf{r},V)$ maps of defect-free regions (Fig. 4e), and STM topographs of multiple samples (Supplementary Figure 2). We also note that the relative amplitudes of CDW ordering wave vectors do not change with magnetic field reversal in the cousin compound $CsV_3Sb_5$ (Supplementary Figure 3).

Our experiments reveal a pronounced rotation symmetry breaking in the $2a_0$ x $2a_0$ CDW phase of $KV_3Sb_5$. Given the reports of unidirectional CDWs [24] and anisotropic transport measurements [31,45] in $CsV_3Sb_5$, which clearly break the rotation symmetry of the lattice, our experiments establish rotation symmetry breaking as a unifying feature of this family of kagome superconductors. A potential explanation for the rotation symmetry breaking in the $2a_0$ by $2a_0$ CDW phase may lie in the coupling between adjacent kagome layers stacked along the $c$-axis, in which the CDW is offset by a lattice vector [42]. This would in principle generate a preferred direction that is seen in our data, but more experiments and theoretical modeling will be necessary to confirm this. Despite a close resemblance between electronic band structures of $KV_3Sb_5$ and $CsV_3Sb_5$, an intriguing difference between the two systems is that $KV_3Sb_5$ does not show the $4a_0$ charge ordering. It may be possible that small differences in the chemical potential between the two materials play a large role, and that the $4a_0$ charge order vanishes with doping, in analogy to the $4a_0$ order in cuprates [46]. Our measurements also highlight the difficulty of determining the morphology of the CDW phases – "star of David" versus inverse star of David [39,40] – exclusively from STM data, since inherent susceptibility of $AV_3Sb_5$ towards rotation symmetry breaking [41] makes the interpretation of the spatial signature in d$I$/d$V(\mathbf{r},V)$ maps challenging.

Contrasting observations reported here and those in Ref. 23 regarding the dependence of the CDW on external magnetic field pose an intriguing question regarding the dichotomy of the results. We note that subtle STM "tip changes" can dramatically change the apparent CDW amplitudes (Fig. S4). Ruling out these trivial measurement artifacts leading to the substantial CDW amplitude change, it may be conceivable that small chemical potential variations between different areas of nominally the same crystals could explain the discrepant results. However, based on our experiments that show the absence of CDW tunability by field direction in all samples and crystals where measurements were performed (5 regions spanning 2 different samples), we can conclude that magnetic field CDW tunability [23] is not a robust feature. It can perhaps be sample and measurement condition dependent. It remains to be seen whether the time-reversal symmetry in $AV_3Sb_5$ is indeed broken by the orbital currents [38,40]. Future experiments using high-resolution spin-polarized STM [47] or neutron scattering [48] could potentially shed light on this issue by measuring any subtle, underlying magnetic background.

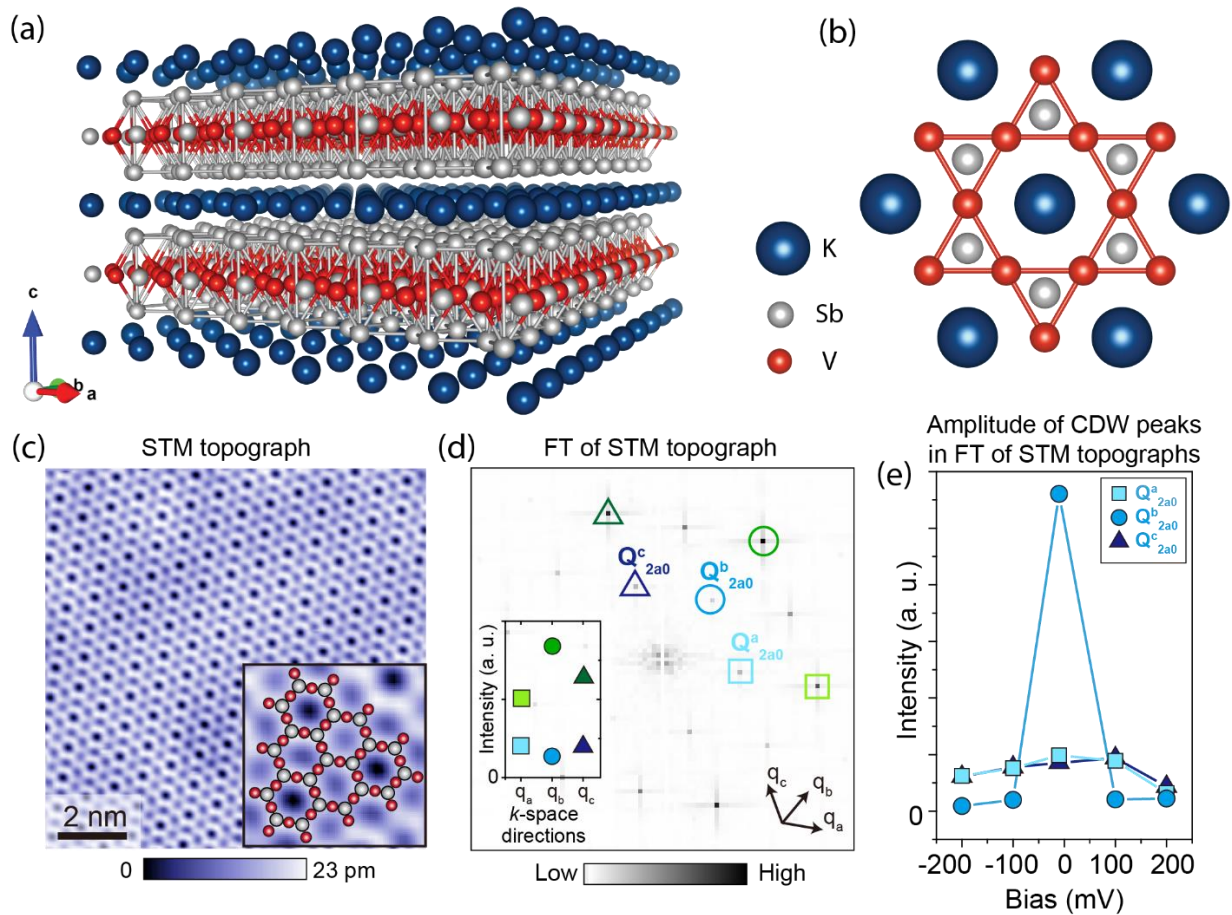

**Figure 1. Crystal structure of $KV_3Sb_5$ and surface morphology**. (a) The 3D crystal structure of $KV_3Sb_5$ depicting two layers, each with one unit cell thickness, stacked on top of one another. (b) Atomic structure of different *ab*-planes superimposed on top of one another: triangular K layer, hexagonal Sb layer and the kagome V-Sb slab. (c) STM topograph of approximately 11 nm square region of the Sb surface, and (d) its associated Fourier transform. Inset in (c) shows a zoom-in on a small region of the topograph, portraying the $2a_0$ intensity variation with the V-Sb kagome layer superimposed on top. The three different CDW peaks and atomic Bragg peaks are enclosed by shapes of different color. Inset in (d) shows the amplitudes (peak heights) of atomic Bragg peaks and the CDW peaks in the topograph in (d). (e) Amplitudes of the three CDW peaks in FTs of five different STM topographs acquired over the same area of the sample with the same STM tip (sample A). STM setup condition: (c) $I_{set}$ = 400 pA, $V_{sample}$ = 40 mV. (e) $I_{set}$ = 400 pA, $V_{sample}$ = 200 mV; $I_{set}$ = 200 pA, $V_{sample}$ = 100 mV; $I_{set}$ = 60 pA, $V_{sample}$ = -10 mV; $I_{set}$ = 200 pA, $V_{sample}$ = -100 mV; $I_{set}$ = 400 pA, $V_{sample}$ = -200 mV. Magnetic field is set to 0 T.

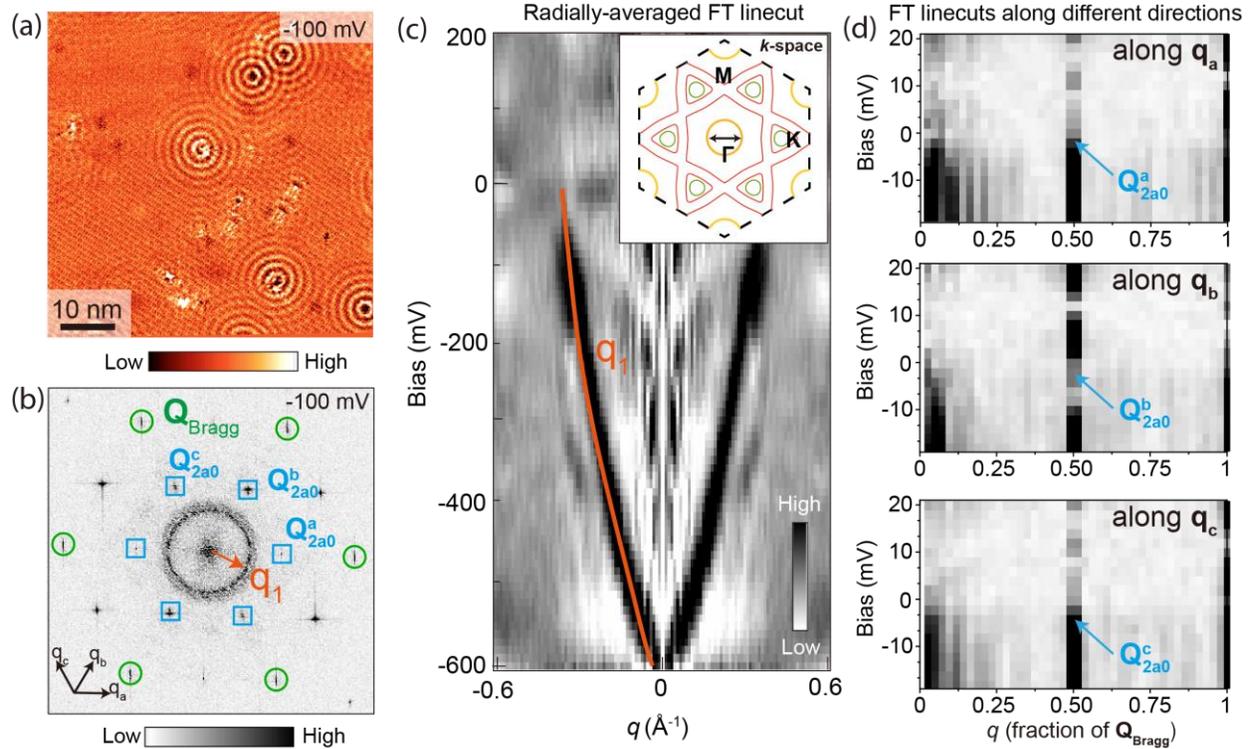

**Figure 2. Spectroscopic-imaging scanning tunneling microscopy of the Sb surface.** (a) Representative d$I$/d$V$(**r**,V) map of approximately 60 nm square region of the Sb termination at -100 mV (sample A), and (b) its associated Fourier transform (FT). The prominent real-space modulations in (a) centered around impurities correspond to the reciprocal wave vector **q$_1$** in (b). Green circles (blue squares) in (b) denote the atomic Bragg peaks (CDW peaks). (c) Radially-averaged line cut of FTs of normalized d$I$/d$V$(**r**,V)/($I$(**r**,V)/V) maps as a function of bias, acquired over region in (a). Orange line in (c) approximately denotes the dispersion of **q$_1$**. Inset in (c) is a schematic representation of the Fermi surface, and the black arrow near the $\Gamma$ point denotes a scattering vector related to **q$_1$** observed in spectroscopic maps. (d) FT linecuts of d$I$/d$V$(**r**, $V$) maps along three lattice directions as a function of bias, acquired over a 12 nm area at ~4.5 K. STM setup conditions: (a) $I_{set}$ = 200 pA, $V_{sample}$ = -100 mV, $V_{exc}$ = 10 meV, B = 0 T; (c) $I_{set}$ = 900 pA, $V_{sample}$ = 300 mV, $V_{exc}$ = 10 meV, B = 0 T. (d) $I_{set}$ = 400 pA, $V_{sample}$ = 20 mV, $V_{exc}$ = 1 meV, B = 0 T.

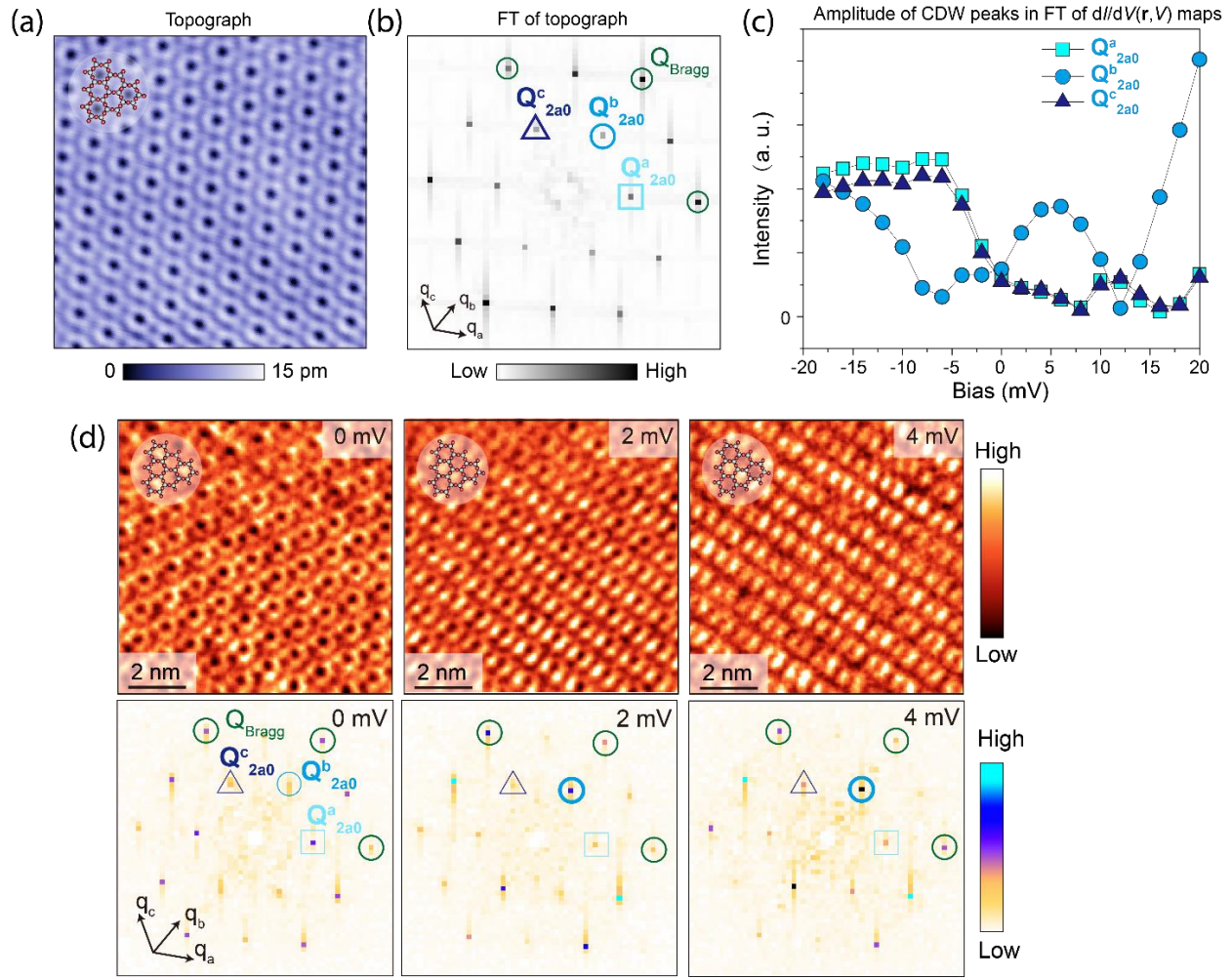

**Figure 3. The anisotropy between inequivalent charge density wave (CDW) directions and its atomic-scale signature in spectroscopic maps.** (a) STM topograph of approximately 10 nm square Sb surface (sample A), and (b) its associated Fourier transform (FT). A layout of the V-Sb atomic structure is marked in the upper left corner of (a) (V is denoted by red spheres and Sb by gray spheres). (c) The amplitude of different CDW peaks in the FTs of d$I$/d$V$(**r**,$V$) maps as a function of STM bias $V$. While the dispersion of $\mathbf{Q}^a_{2a0}$ and $\mathbf{Q}^c_{2a0}$ is nearly identical, the dispersion of $\mathbf{Q}^b_{2a0}$ is markedly different. (d) High-resolution d$I$/d$V$(**r**,$V$) maps at 0 mV, 2 mV, and 4 mV over the region in (a) (top row) and their associated FTs (bottom row). The three CDW peaks are enclosed by a triangle, a circle and a square marker in all panels, using different shades of blue. STM setup conditions: (a) $I_{set}$ = 150 pA, $V_{sample}$ = 10 mV, $V_{exc}$ = 10 meV, $B$ = 0 T; (c) $I_{set}$ = 400 pA, $V_{sample}$ = 20 mV, $V_{exc}$ = 1 meV, $B$ = 0 T; (d) $I_{set}$ = 150 pA, $V_{sample}$ = 10 mV, $V_{exc}$ = 1 meV, $B$ = 0 T.

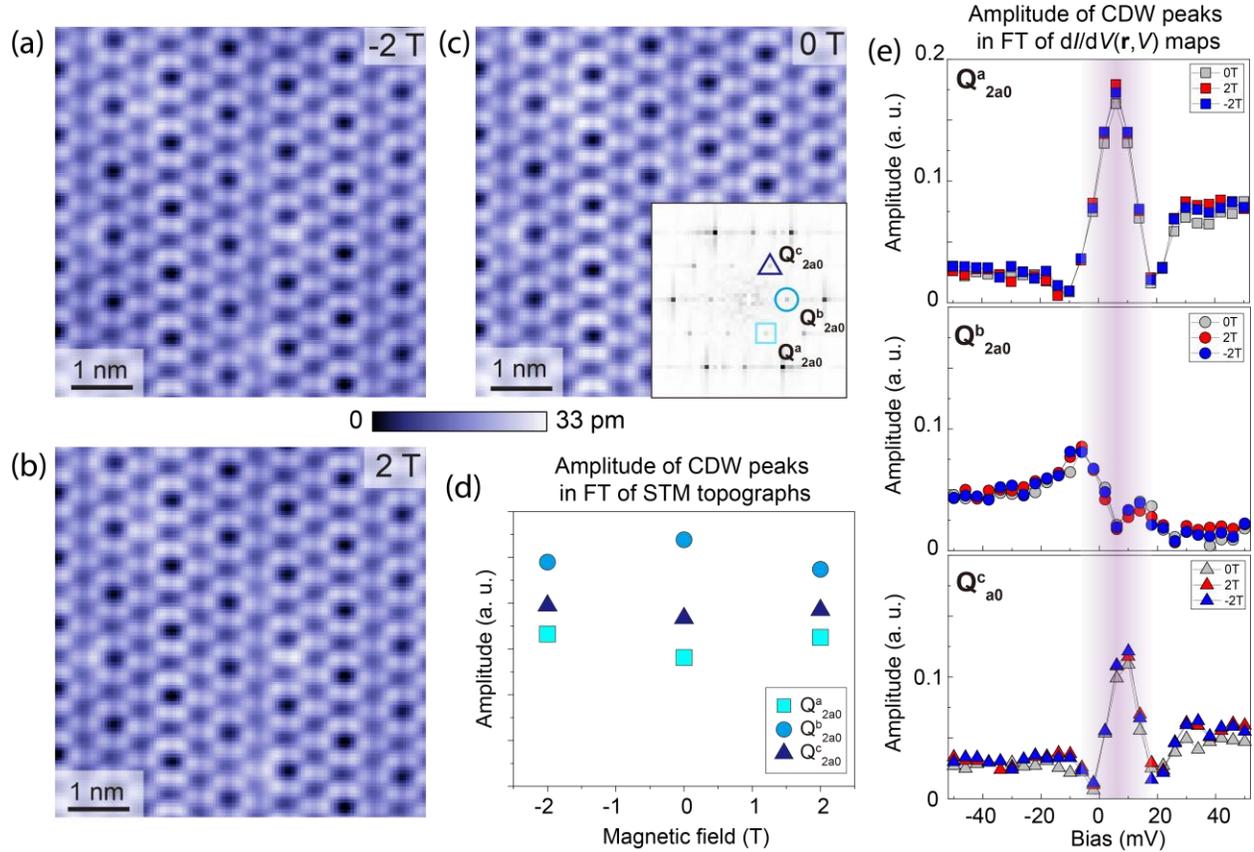

**Figure 4. Insensitivity of the charge density wave (CDW) to external magnetic field direction.** (a-c) STM topographs of an identical region of Sb surface acquired at different magnetic fields (sample B). Inset in (c) shows the Fourier transform (FTs) of the topograph in (c), with the three CDW peaks denoted by different symbols. (d) Amplitudes of the three CDW peaks in FTs of STM topographs in (a-c) acquired at different magnetic field B applied perpendicular to the sample surface. The sign associated with B denotes the reversal of magnetic field direction (parallel vs antiparallel to $c$-axis). (e) Amplitudes of CDW peaks in d$I$/d$V$($\mathbf{r}$,$V$) maps as a function of magnetic field and bias $V$, acquired over the region in (a-c). The diffuse vertical purple line denotes the approximate energy range where the peak in the dispersion related to $\mathbf{Q}^a_{2a0}$ and $\mathbf{Q}^c_{2a0}$ coincides with the minimum in $\mathbf{Q}^b_{2a0}$ dispersion. STM setup conditions: (a-d) $I_{set}$ = 100 pA, $V_{sample}$ = 50 mV; (e) $I_{set}$ = 100 pA, $V_{sample}$ = 50 mV, $V_{exc}$= 4 mV.

**Methods**

Single crystals of $KV_3Sb_5$ were grown and characterized as described in more detail in Ref. [33]. We cold-cleaved and studied five different $KV_3Sb_5$ crystals, all of which exhibited qualitatively the same phenomena described in the main text. STM data was acquired using a customized Unisoku USM1300 microscope at approximately 4.5 K. Spectroscopic measurements were made using a standard lock-in technique with 915 Hz frequency and bias excitation as also detailed in figure captions. STM tips used were home-made chemically-etched tungsten tips, annealed in UHV to bright orange color prior to STM experiments.

**Acknowledgements**

I.Z. gratefully acknowledges the support from the National Science Foundation grant NSF-DMR-1654041 and Boston College startup. S.DW., and B.R.O., and T.P. acknowledge support via the UC Santa Barbara NSF Quantum Foundry funded via the Q-AMASE-i program under award DMR-1906325. Z.W. acknowledges the support of U.S. Department of Energy, Basic Energy Sciences Grant No. DE-FG02-99ER45747. L.B. is supported by the NSF CMMT program under Grant No. DMR-1818533. M.Y. is supported in part by the Gordon and Betty Moore Foundation through Grant GBMF8690 to UCSB and by the National Science Foundation under Grant No. NSF PHY-1748958. T.P. was supported by the National Science Foundation through Enabling Quantum Leap: Convergent Accelerated Discovery Foundries for Quantum Materials Science, Engineering and Information (Q-AMASE-i) award number DMR-1906325.